\documentclass[preprint]{aastex631}

\usepackage{amsmath}
\usepackage{graphicx} 

\begin{document}

\title{Characterization of the Deep, Extended Kuiper Belt in the Galactic Disk}

\author[0000-0001-8821-5927]{Susan D. Benecchi}
\affiliation{Planetary Science Institute, 1700 East Fort Lowell, Suite 106, Tucson, AZ 85719}
\date{May 2024}

\author[0000-0003-0333-6055]{Simon B. Porter}
\affiliation{Southwest Research Institute, 1301 Walnut St., Suite 400, Boulder, CO 80302, USA}

\author[0000-0002-3323-9304]{Anne J. Verbiscer}
\affiliation{University of Virginia, P.O. Box 400325, Charlottesville, VA 22904-4325, USA}
\affiliation{Southwest Research Institute, 1301 Walnut St., Suite 400, Boulder, CO 80302, USA}

\author[0000-0001-6942-2736]{David W. Gerdes}
\affiliation{Department of Physics, University of Michigan, Ann Arbor, MI 48109, USA}
\affiliation{Department of Astronomy, University of Michigan, Ann Arbor, MI 48109, USA}

\author[0000-0001-6680-6558]{Wesley C. Fraser}
\affiliation{National Research Council of Canada, Herzberg Astronomy and Astrophysics Research Centre, 5071 W. Saanich Rd. Victoria, BC, V9E 2E7, Canada}
\affiliation{Department of Physics and Astronomy, University of Victoria, Elliott Building, 3800 Finnerty Road, Victoria, BC V8P 5C2, Canada}

\author[0000-0002-9179-8323]{Lowell Peltier}
\affiliation{National Research Council of Canada, Herzberg Astronomy and Astrophysics Research Centre, 5071 W. Saanich Rd. Victoria, BC, V9E 2E7, Canada}
\affiliation{Department of Physics and Astronomy, University of Victoria, Elliott Building, 3800 Finnerty Road, Victoria, BC V8P 5C2, Canada}

\author[0000-0001-7032-5255]{JJ Kavelaars}
\affiliation{National Research Council of Canada, Herzberg Astronomy and Astrophysics Research Centre, 5071 W. Saanich Rd. Victoria, BC, V9E 2E7, Canada}
\affiliation{Department of Physics and Astronomy, University of Victoria, Elliott Building, 3800 Finnerty Road, Victoria, BC V8P 5C2, Canada}

\author[0000-0003-0854-745X]{Marc W. Buie}
\affiliation{Southwest Research Institute, 1301 Walnut St., Suite 400, Boulder, CO 80302, USA}

\author[0000-0001-5018-7537]{S. Alan Stern}
\affiliation{Southwest Research Institute, 1301 Walnut St., Suite 400, Boulder, CO 80302, USA}

\author{Kelsi Singer}
\affiliation{Southwest Research Institute, 1301 Walnut St., Suite 400, Boulder, CO 80302, USA}

\author[0000-0003-4143-4246]{Tsuyoshi Terai}
\affiliation{Subaru Telescope, National Astronomical Observatory of Japan 650 North A`ohoku Place, Hilo, HI 96720, USA}

\author[0000-0002-0549-9002]{Takashi Ito}
\affiliation{Center for Computational Astrophysics, National Astronomical Observatory of Japan, Osawa 2-21-1, Mitaka, Tokyo, 181-8588, Japan}

\author[0000-0002-3286-911X]{Fumi Yoshida}
\affiliation{University of Occupational and Environmental Health, 1-1 Iseigaoka, Yahata, Kitakyushu 807-8555, Japan}
\affiliation{Planetary Exploration Research Center, Chiba Institute of Technology, 2-17-1 Tsudanuma, Narashino, Chiba 275-0016, Japan}

\author[0000-0003-1080-9770]{Darin Ragozzine}
\affiliation{Department of Physics and Astronomy, Brigham Young University, Provo, UT 84602, USA}

\author{Bryan Holler}
\affiliation{Space Telescope Science Institute, Baltimore, MD, USA}

\begin{abstract}

We propose a Roman Space Telescope survey to investigate fundamental properties of the distant solar system in the region of the Kuiper Belt where object characteristics and the size distribution are inaccessible from any other telescope. Our pointing is coincident with the search space accessible to NASA's New Horizons spacecraft meaning, that a discovered object sufficiently near the orbit of New Horizons would potentially be investigated by a close flyby. In addition, numerous objects expected to be discovered by this search can be observed in the distance by New Horizons allowing their surface properties and satellite systems to both be probed. As designed, this survey will discover and determine orbits for as many as 900 Kuiper Belt objects (KBOs), providing a unique opportunity for ground-breaking Kuiper Belt science. It will simultaneously: (1) Probe and characterize the deep Kuiper Belt by identifying objects as small as a few km and taking our understanding of the size distribution to a new level. This has implications for understanding the the standard model (the Streaming Instability) of KBO formation and elucidating crater formation physics on these icy bodies. (2) Open KBO rotation studies, in particular of those objects with long rotation periods, (3) Discover and characterize KBO binaries at large distances, important because their duplicity offers information about object densities at these distant locations from the Sun. (4) Shed light on the cratering history of KBOs and  improving the dating of the surfaces of Arrokoth,  Pluto and Charon in addition to helping to place the 32 distant KBOs New Horizons has observed in context. This project also has synergies with transiting exoplanet studies due to the stellar density of our search fields. Coupled with our timing requirements it is sensitive to discovery of hot Jupiters and hot Neptunes.

\end{abstract}

\section{The Kuiper Belt}
As a result of numerous studies in the past two decades, disks and exozodiacal are found around a significant fraction of forming Sun-like stars \citep{Kral2017}, and exoplanets are abundant \citep[e.g.][]{Yang2024, Ivshina2022, Kaltenegger2019}. However, we view these systems from the outside without a much detail. We are inside our own Solar System enabling us to gain insight to the elements and components which make up the planets, small bodies and dust. From these observations, models are then constructed for our own system with application to external ones \citep{Chambers2023}. The small bodies are particularly fundamental for constructing and constraining these models. They reside throughout our Solar System and their characteristics map, in part, the history of our Solar nebula. Kuiper Belt objects (KBOs) in the outer reaches of the Solar System, their locations and physical properties \textemdash size, population statistics, composition, shape, binarity \textemdash all provide clues to the radial construction of the primordial Solar nebula with respect to materials, thermal history and motion of the giant planets \citep{Nesvorny2021, Nesvorny2023PSJ}. The Kuiper belt has been characterized to have both a ``cold classical'', low inclination/low eccentricity component, and a variety of excited populations including mean motion resonances with Neptune, hot classical, scattered and detached objects spanning a full range of phase space. Characterizing these populations both statistically, and objects within them uniquely, provide key constraints for formation models, which have implications for the smoothness of Neptune's migration, compositional distribution of the Solar nebula and accretion processes in the Solar System \citep[e.g.][]{Morbidelli2007, McKinnon2020}.

Wide-area ground-based surveys have systematically reached a magnitude of 26.5 \citep{Bannister2018, Fraser2024LPI, Yoshida2024} and the coming Vera Rubin Observatory Large Synoptic Survey is projected to reach 28.5 in some deep fields with the development of multi-night stacking techniques $\sim$200 hours of integration time \citet{lsstsciencecollaboration2009lsst}. The Nancy Grace Roman Space Telescope is far more efficient, reaching this depth in about an hour using its wide F146 filter and stable PSF. In 42.3 hours, Roman can reach magnitude 30.5, a critical depth for finding objects in the outer reaches of the Kuiper Belt (Figure $\ref{fig-Hmag}$). Likewise, New Horizons will only be in the Kuiper belt and operational for another $\sim$ 15 years, so if we happen to find an object near its trajectory, it is perfectly timed for fully exploiting the use of the spacecraft in an area of the Solar System that practically speaking will not be visited again in our lifetimes. 
In conclusion, this provides a golden opportunity for the Roman Galactic Disk survey to both probe the disk and to yield fundamental new insights into our outer solar system by probing the Kuiper Belt to much fainter limits than previously possible while also providing objects for New Horizons to study.

\subsection{Impact of NASA's New Horizons Mission}
NASA's New Horizons Mission was specifically designed to characterize the small bodies in the outer reaches of our Solar System and it has made critical unique contributions including: (1) Understanding the Pluto-Charon system at an unprecedented level (e.g. \citet{Stern2015}, \citet{DESCH2017}, \citet{WONG2017}, \citet{KRASNOPOLSKY2020}). (2) Transformation of our understanding of small KBOs yielding new constraints on chemistry \citep{Grundy2020}, cratering \citep{Singer2019}, and processes associated with planetesimal accretion \citep[e.g.][]{Nesvorny2021, Nesvorny2022, Nesvorny2023, Nesvorny2023PSJ, Stern2023}. (3) It also allowed us to make comparisons of small bodies, in particular placing small KBOs in context with the giant planet icy satellites and inter-KBO dynamical populations \citep{Verbiscer2019,Verbiscer2022}. It continues to be healthy and moving through the deep Kuiper Belt. And like all missions, New Horizons has settled some science questions and opened others. Although the probabilities are low, the flyby of an object 2-3 times further from the Sun than Arrokoth would be completely unique, providing a revolutionary perspective and the only one plausible in our lifetimes. Independent of this opportunity, however, this proposed survey will illuminate fundamental details of this region of space which have never been feasible for us to consider investigating and which nicely compliment the mission.

\section{Kuiper Belt and Astrophysics Science Enabled by the Galactic Disk Survey}

\subsection{Kuiper Belt Science}
\begin{center}
\begin{figure}
\includegraphics[scale=0.55]{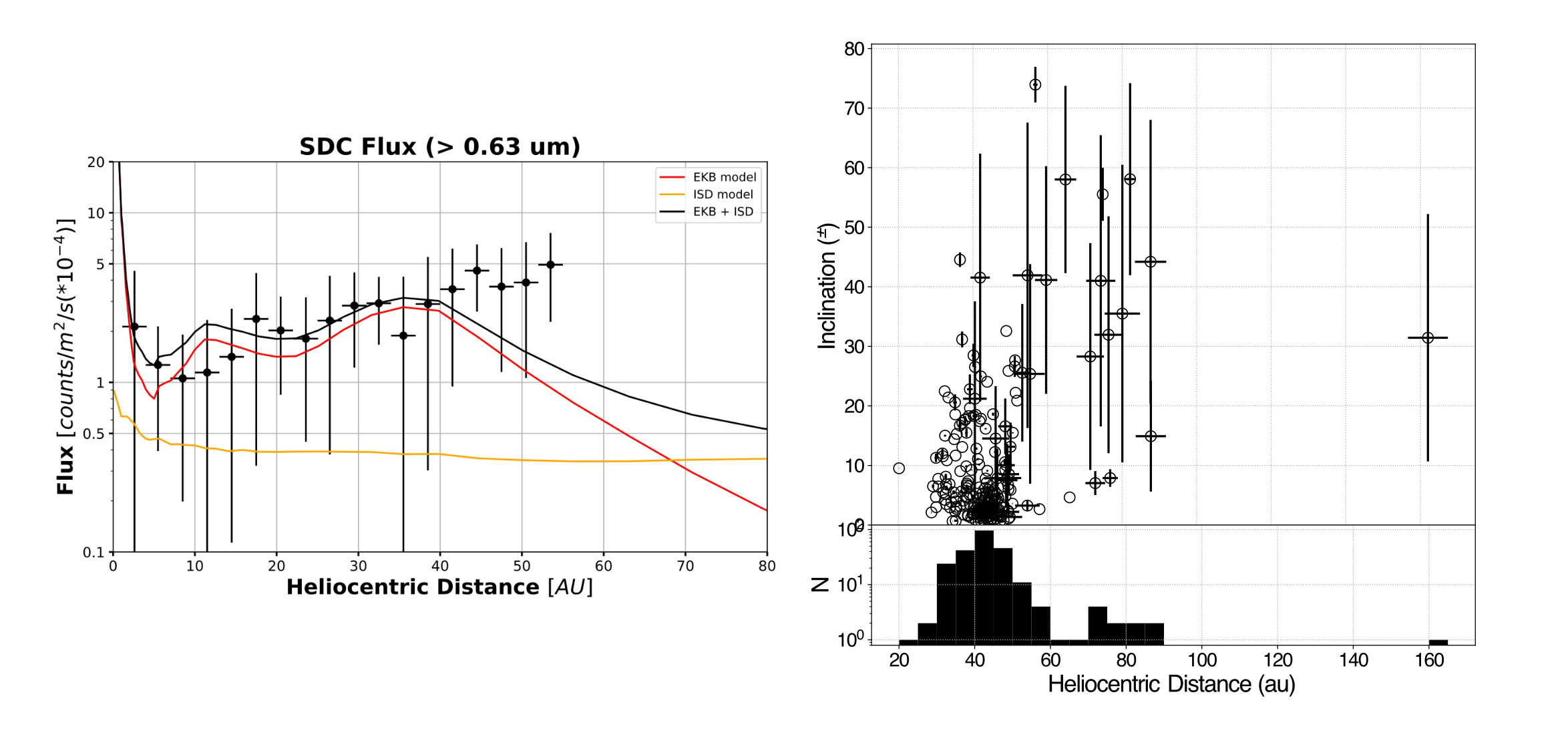}
    \caption{(left) Figure 4 from \citet{Doner2024}, SDC flux estimates for particles with a radius greater than 0.63 $\micron $ from 1 to 55 heliocentric astronomical units. Each point is an average of the flux measured by each film across each 3 au traversed by the New Horizons spacecraft. (right) Figure 4 from \citet{Fraser2024LPI}, showing the inclination and distance of the New Horizons discoveries from Subaru observations 2020-2023. A histogram of the heliocentric distances are shown in the bottom panel, with bin width of 5 au chosen to be similar to the typical uncertainty of the objects with best-fit heliocentric distance R $>$ 70 au. Both of these figures indicate that ojects and material continue to be present past our current mapping of the Kuiper Belt. }
\label{fig-KBOsPlot}
\end{figure}
\end{center}

Due to its depth ($m \sim 30.5$) and pointing near the ecliptic plane (ecliptic latitude of $\sim 2$ degrees) and the wide field nature of the RST FOV we will, for the first time, characterize any deep, extended Kuiper Belt in detail. Current ground-based detections of distant KBOs (Figure \ref{fig-KBOsPlot}; \citep{Fraser2024LPI,Yoshida2024}), as well as measurements by the New Horizons Student Dust Counter \citep{Doner2024}, give evidence for a potentially sizable component to the disk further out. The $\sim 960$ KBOs (Table \ref{tab:timetable}) that our proposed Roman Space Telescope galactic disk project can discover in a 3-field footprint  will include a significant fraction of small objects with magnitudes fainter than 28 (Figure $\ref{fig-Hmag}$ a region of the size distribution that is almost totally unexplored. Only one KBO with $m>28$ has ever been recorded in the literature \citep{Bernstein2004}. The Cycle 1 JWST deep survey (GO-1191 P.I. Stansberry), which covers a  footprint of 0.02 degrees squared (14$\%$ of the Roman field), is projected to go to 29.5. We propose a 3-Roman field survey to go an order of magnitude fainter providing a statistically significant sample of KBOs in a new size regime to reveal fundamental properties of our outer solar system. \textbf{Likewise, with only $\sim32$ days of total integration time this project discovers and characterizes approximately the same number of objects as many 4-year ground-based surveys of an entirely new population of objects.}

\begin{center}
\includegraphics[scale=0.5]{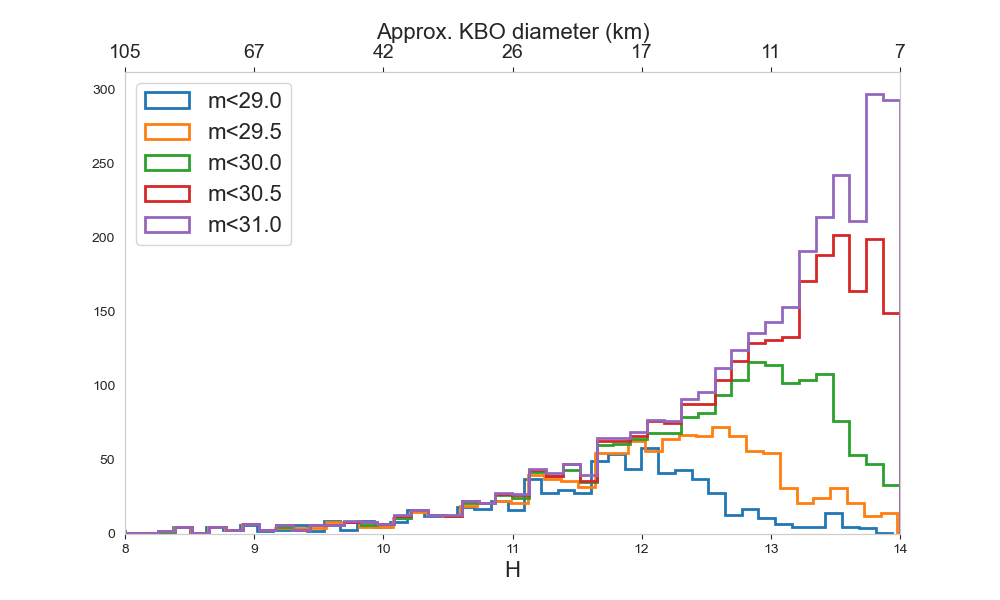}
\figcaption{\label{fig-Hmag}
H distribution for KBOs discoverable at the NH search location as a function of survey depth. The corresponding size of these objects, assuming an albedo of 10\%, is shown along the top axis. The smallest KBOs accessible to ground-based surveys have sizes $\gtrsim 30$~km.
}
\end{center}

\subsection{The Distant Kuiper Belt}
Fundamentally, this survey will elucidate the distant or deep Kuiper belt in ways unobtainable otherwise \citep{Kavelaars2020}. If the size frequency distribution (SFD) of the distant bodies follows the slope of alpha $\sim~0.3$ for normal KBOs in a similar size range, the density of these objects at r$=$30 mag will be roughly 11x higher than at the current state of the art Subaru depths. As such, considering the Roman areal coverage, we would expect to detect $\sim 60$ newly detected objects at 70 au (compared to the Subaru survey's 11, Fraser et al. 2024 submitted). They would also be far more robust than the Subaru detections because the observation spacing within a season and covering two seasons will result in robust orbit measurements from which their origins could be evaluated. If these distant KBOs originated with the rest of the normal dynamically excited KBOs, then we would expect them to have moderate to high inclinations and eccentricities, and must be coupled to Neptune (be in a resonance). It they originated from a cold disk and were pushed out during smooth migration, then they either have originated with the cold classicals and be found with low-inclinations and moderate eccentricities (and red colours if we could measure them), still dynamically coupled to Neptune. Or, if they are in fact a second Kuiper Belt, they would be entirely uncoupled from Neptune, though predictions of their orbital distribution are hard to make with current data.

\subsection{Size Distribution, KBO Formation \& Cratering} 
KBOs, in particular, may be less processed by later evolutionary forces (e.g., collisional evolution, surface geology, outgassing), and thus provide a unique window into the planetesimal formation era.  Predictions from competing hypotheses about planetesimal formation diverge for smaller bodies (less than ~ 1-10 km diameter), where hierarchical accretion followed by collisional evolution predicts many small bodies \citep[e.g.][]{Schlichting2013, Bottke2023PSJ}, and models of streaming instabilities/gravitational collapse would predict fewer small bodies \citep[e.g.][]{Youdin2007, Chiang2010AREPS, Nesvorny2010, Johansen2015, Davidsson2016, Simon2017, Abod2019}.  New Horizons added a new piece to the puzzle, when it discovered that crater size-frequency distributions Pluto, Charon, and Arrokoth showed a deficit of smaller objects \citep{Singer2019,Spencer2020}.  

This deficit of small objects was seen in the crater populations for impactor/KBO sizes of ~200 m to 1 km, but it is unclear what happens below this size.  Additionally, observing the extant KBO populations rather than scaling from craters would provide an unprecidented constraint on Solar System formation scenarios, including properties such as the primordial pebble size, the accretion efficiency, and the masses of the initial clouds required to produce planetesimals.  100-m objects have effective magnitudes of r$\sim 29$ assuming $12\%$ albedos which is at the absolute detection limit of even the JWST pencil beam survey \citet{Stansberry2021}. To confidently detect and characterize the small KBO Science Frequency Distribution (SFD) requires detecting objects well below with diameters $\sim$100m, impossible from anything other than the proposed Roman survey (Figure \ref{fig-SFD}).

Additionally, with the KBO and crater SFDs in hand, one can directly measure the age of the surfaces on which the craters are superimposed. The impactor flux is given by the known orbital distribution of the impactors so the number of craters/impactor rate provides the age since the last surface refresh (eg. nitrogen ice turnover, or sublimation erosion). Knowing these ages can put a lower limit on some of Pluto's and Charon's surfaces. If they are much younger than the Pluto-Charon forming event, that would provide extremely powerful knowledge regarding the cosmogony of the Outer Solar System. For example, with knowing the ages of the oldest surfaces on Pluto and Arrokoth would allow us to constrain the ages of each body. For Pluto, this would provide an age of the Pluto-Charon forming impact and for Arrokoth, this would mark the age of formation of actual primordial bodies. Our current estimates are drawn from assumption-dependent age estimates based on dynamical models.

\begin{center}
\includegraphics[scale=0.15]{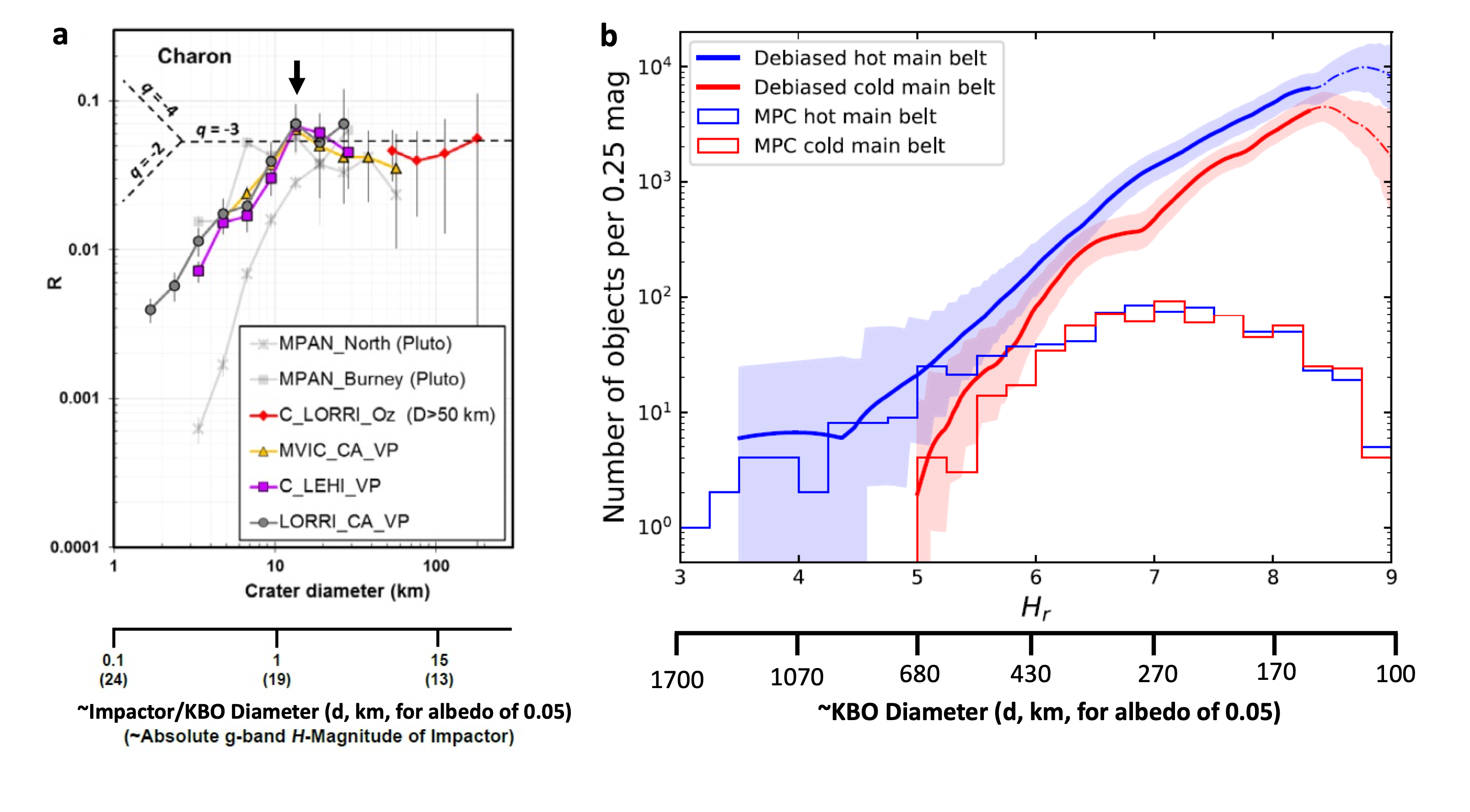}
\figcaption{\label{fig-SFD}
(a) Crater distributions on Charon’s geological units (shown in a traditional R-plot style), illustrating the break (arrow) to a shallower size-frequency distribution slope \citep[modified from][]{Singer2019} for craters less than ~10 km or impactors/KBOs less than ~1 km.  Reference differential size-frequency distribution slopes are shown with dashed lines.  (b) Telescopic survey results for larger KBOs showing the deviations from the currently observed KBOs (MPC histograms) compared to the modeled debiased populations (solid curves) \citep[from][]{Petit2023}.  The debiasing is based on models of the survey efficiencies and filling in the smaller sizes with actual observations would be highly constraining.
}
\end{center}

\subsection{Physical Characteristics: Rotation Curves \& Binaries}
Likewise, because Roman is in space, its observations are not limited to the 24-hour rotation cycle. This enables significant work related to lightcurve measurements for any moving objects in the survey which have long periods. Work by \citet{Kecskemethy2023} using the K2 dataset find a significant fraction of KBOs to have periods longer than $\sim20$ hours. This result is also supported by New Horizons observations of distant KBOs \citep{Porter2022,Porter2023} and uniquely enabled by Roman observations. Understanding how the rotation properties (amplitudes and periods) change as a function of size will provide unique insights into planetesimal formation. Observations to reach the required search depth will by default provide a baseline that covers nearly twice the longer period peak distribution for brighter KBOs. Short period objects will also be observed, but the overall statistics will be far more significant for the former. We will also be sensitive to any potential activity, although it is not expected. 

The high-resolution of Roman also enables the discovery and characterization of close binaries. When observed with HST, about 10\% of objects are well-resolved binaries and about 10\% of objects are blended but resolvable with Point Spread Function (PSF) fitting which would be possible for these proposed observations (Porter et al., in prep.). Observations at multiple epochs also improves binary discovery and allows for approximate orbital fits. 

The fraction of binaries at these small sizes is almost completely unknown and is an important probe of planet formation. There is some evidence from binary cratering on Pluto that there are small binaries. The proposed survey would provide unique and crucial insights into this question. One anticipated challenge in looking for binaries around smaller KBOs is that these objects may naturally have smaller separations. Suppose that the typical separations scale with Hill radius \citep[comparable to known binaries;][]{Grundy2019}. Since we are probing a population that is typically $\sim$10 times smaller, instead of typical angular separations of tenths of an arcsecond, typical separations may be hundredths of an arcseconds or a fraction of a Roman pixel. Under this hypothesis, we would expect that only the $\sim$2\% of known objects with rather wide separations would be detectable as binaries in this survey. At our deepest search depths, the expected discovery of $\sim$20 binaries would allow our survey to confirm or refute the hypothesis at the $\sim$3-$\sigma$ level that the binary fraction at wide separations continues to smaller objects with a similar Hill radius distribution. Likewise, by measuring lightcurves and searching for binaries, we can provide direct observational constraints on the precursor comet population.

The Roman database will be a treasure trove for better understanding the rotational and binary properties of this population which is particularly helpful because binaries by nature give us a way to discern system mass and by extension object densities.

\begin{center}
\includegraphics[scale=0.35]{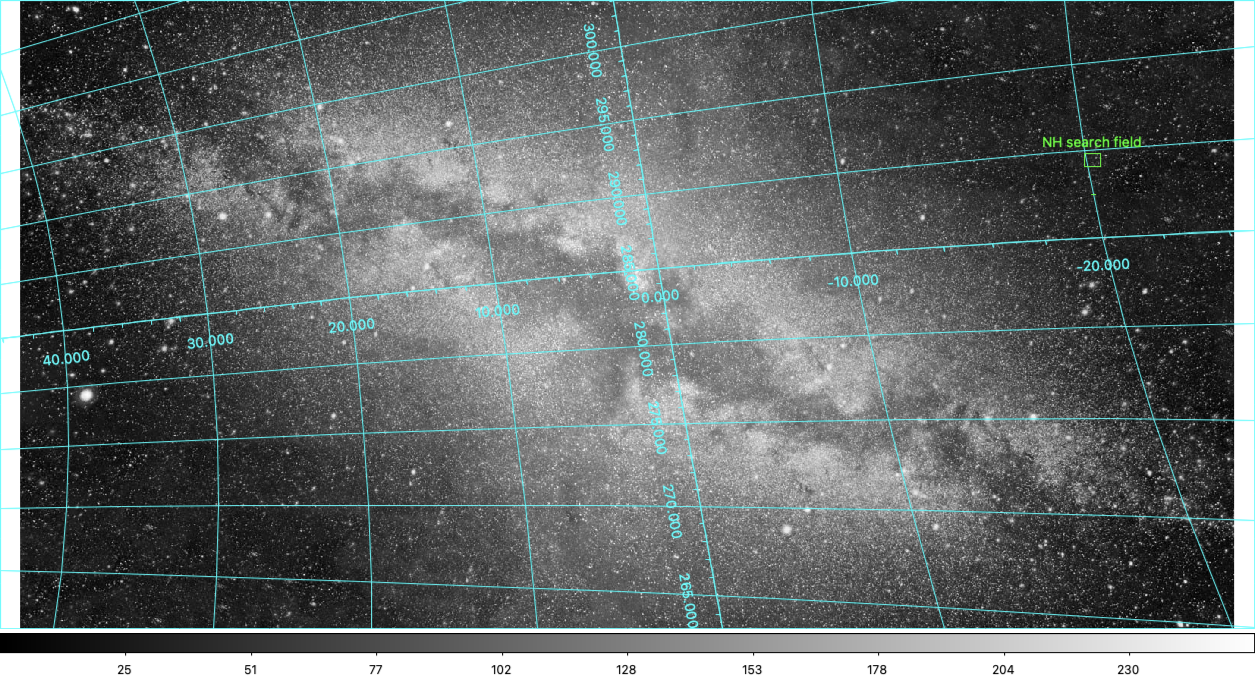}
\figcaption{\label{fig-image}
Placement of the New Horizons search field (green square) relative to the galactic plane. 
}
\end{center}

\section{Roman Survey Requirements}

The Roman project has four highly designed surveys, two related to the Galaxy and two for observations at high latitudes. Most astrophysics projects find objects moving through their fields an annoyance because they temporarily intersect with, or land near, fixed objects. These objects are treasures for Solar System observers. Because the field where New Horizons will fly is close to the Galactic Plane, (Figure \ref{fig-image}) we propose a three-field survey directed in over this region as part of the Galactic Plane Survey. This provides a unique opportunity to both: (1) probe the size space in the Kuiper Belt beyond which we can access from the ground and (2) potentially find a flyby object for New Horizons to visit before its exit from this region of the Solar System. We list in this section the requirements for our search and the areas of flexibility within them. 

Observations are to be obtained in the F146 filter with 3 pointings near RA=289.4, DEC=-20.2 (Figure \ref{fig-RomanField}). This field is defined as the search region for objects that are at least 66 au from the Sun between January 1, 2027 and 2040, and lying within 7 degrees on the sky of New Horizons' position on that date as derived from the OSSOS$++$ Kuiper Belt model (\citet{Petit2023}). All these objects pass within 1 au of New Horizons and have $V <30$ in mid-2027. Based on this model we could expect to discover $\sim13$ objects that approach New Horizons within 1 au in addition to the $\sim 900$ Survey KBOs. While the flyby probabilities are low, they are not non-existent, in particular if the mostly unsampled 5:2 resonant population is large, this number could increase by a factor of a few.

The field needs to be observed 3x per season for 2 seasons: (1) near the beginning of the observability window [which is about 7 weeks long: April 4 - May 25 and August 28 - October 16],  (2) near the center of the window and (3) near the end of the observing window. Each window requires 42.3 hours per Roman field with slight adjustments/telescope nudges to the pointing at intervals during these timespans to account for the non-sideral motion of the KBOs (this could be done by repeat visits of $<10$ hours in equal duration at the 55 second per exposure cadence described on the Roman website (${https://roman.gsfc.nasa.gov/science/WFI_technical.html}$) accumulating to 42.3 hours per Field, with the pointing adjusted to match approximate non-sidereal motion, or consecutive 1 hour windows where we shift the telescope pointing a bit between each hour for the 43.2 contiguous hours). This allows us to use a variety of stacking groupings to reach varying depths down to the optimal depth and also to link object discoveries over time baselines which allow their orbits to be determined to enough precision (a few arcseconds) for JWST or spacecraft targeting. For the optimal search to $m=30.5$ we require 761.4 total hours of integration. Each individual visit is a duration of 126.9 hours, each epoch (3x the individual visit) sums to 380.7 hours and the total search covering both epochs (6 visits to the field total) to 761.4 hours. 

Table \ref{tab:timetable} shows the improvement by depth of the number of KBOs projected to be discovered total and, within 1 au of the spacecraft. The optimal search takes us to  $m=30.5$ and requires 761.4 hours. Descopes to these requirements might be collecting data at only 2x per season or decreasing the overall exposure time, but both come at the expense of number of objects discovered or unique orbits acquired. 

\begin{center}
\includegraphics[scale=0.8]{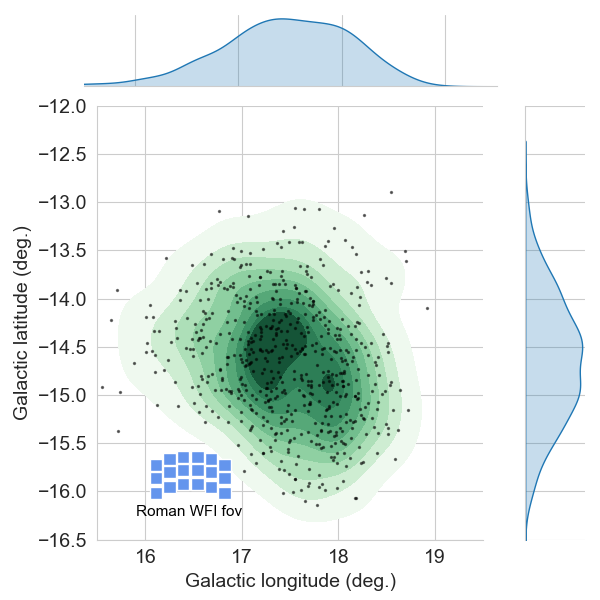}
\figcaption{\label{fig-RomanField}
New Horizons search field with the Roman footprint. This plot illustrates KBOs (black points) that pass within 1 au of New Horizons between 2027-2040 and have $V<30$ in mid-2027, based on the 20$\times$ oversampled extended Kuiper Belt model of Kavelaars et. al (2023). The contours show the concentration of potential objects in a given location. The approximate size and shape of the Roman wide-field instrument focal plane is shown in the bottom left corner for reference.
}
\end{center}

\startlongtable
\begin{deluxetable*}{llllccccccc}
\tabletypesize{\scriptsize}
\tablecaption{Expected KBO for a 3-Roman field of view (0.9 sq.\ deg)}

        \tablehead{
        \colhead{Search Depth} &
        \colhead{Total KBOs} &
        \colhead{1 au KBOs} &
        \colhead{1 Epoch Int. Time} &
        \colhead{2 Epoch Int. Time}\\
        \colhead{Vmag} &
        \colhead{No.} &
        \colhead{No.} &
        \colhead{Hours} &
        \colhead{Hours} &
}
\startdata 
29.0 & 270 & 3.0 &  22.8&   46.5\\ 
29.5 & 450 & 4.5 &  58.5 &  117\\ 
30.0 & 690 & 7.2 &  149.1 & 298.5\\ 
30.5 & 960 & 12.6 & 380.7 & 761.4\\ 
\enddata
\label{tab:timetable}
\end{deluxetable*}


\subsection{Analysis Techniques}

Because Pluto was traversing the galactic plane when New Horizons was launched, most of the search region for flyby targets by the spacecraft have been in highly populated stellar fields. We propose this Roman survey to be carried out in the specified location because it allows us to both characterize the deep Kuiper Belt as well as allow for the probability, albeit low, of one final flyby object for the spacecraft. Over time, special techniques have been developed to cope with these high sky backgrounds and false positives (Figure \ref{fig-sample};\citet{Buie2024} ). Additionally, techniques have been developed to stack sets of short exposures collected close to each other in time to be compared with later sets of close-in-time exposures for tracking these objects and determining their orbits (\citet{Fraser2024LPI}; \citet{Yoshida2024}). These techniques will be applied to the Roman Galactic Plane survey dataset. 

\begin{center}
\includegraphics[scale=0.3]{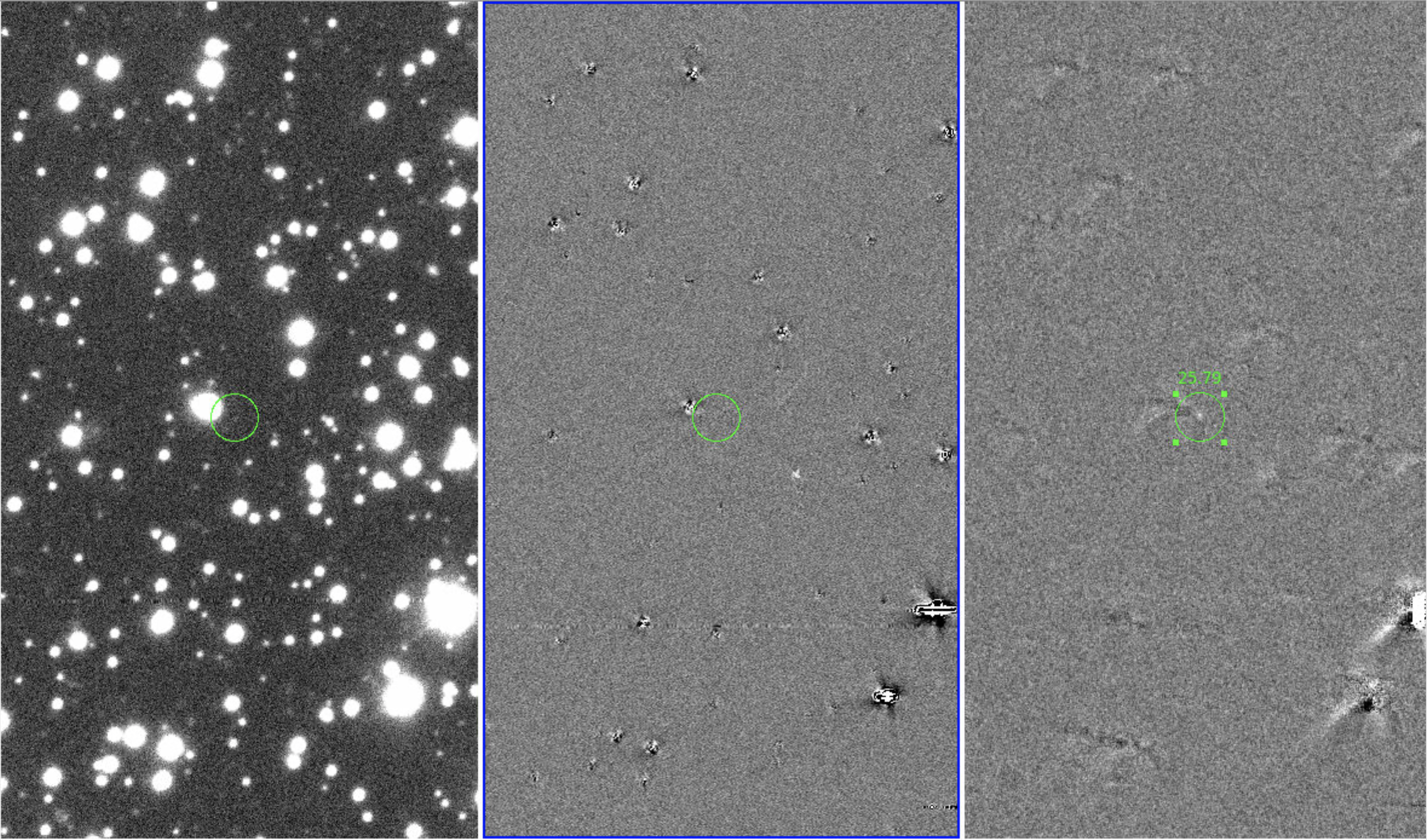}
\figcaption{\label{fig-sample}
Sample of image series stacks followed by subtraction of a sky template of stable sources and then identification of moving objects within those differenced images. 
}
\end{center}

\section{Synergies}
While our focus is on moving objects in the Solar System the basic requirements for our observations support other science cases that benefit from longer duration stares and long time bases such as microlensing calibrations and transiting exoplanets.

Our Galactic Plane Survey as defined is somewhat similar to that of the Multiband Imaging Survey for High-Alpha PlanetS (MISHAPS) which searches for hot Jupiters in the Galactic bulge and their occurrence rates from the ground (\citet{Penny2020}). Without running full-up models which are beyond the scope of this white paper we estaimte that within our 0.84 sq. deg area there would be $\sim$ 380,000 thick disk stars brighter than W146=23 which would nicely separate from the thin disk in a color magnitude diagram. While we don't require a second color for our survey, it might be a requirement for the larger pre-defined Roman Galactic Plane survey. Proper motions for these stars would improve the results. The actual number of detections might be in the low hundreds. The stars in our field would be mostly from the low-metallicity thick disk and halo populations, an interesting region of phase space for exoplanet researchers. The occurrence rate of hot Neptunes is higher than of hot Jupiters, but our dataset would be sensitive to them over a smaller range of magnitudes, so the yield for both might be comparable.

Another impact these observations might have is for microlensing by helping to constraining Galactic models by modeling the star counts – the combination of depth and location is unique. Again, measurements over different epochs are required for measuring proper motions. 

For the New Horizons spacecraft itself the Roman data could enhance the science value by providing disambiguation of multiple sources which lie within a New Horizon pixel which may lead to false positive KBO detections, and  identification of stars which have varied in magnitude or position between the New Horizons and Roman observations.

\section{Summary}
In summary this white paper suggests using 761.4 hours of Roman Galactic Plane Survey time spread over 2 seasons within a few degrees of the Galactic Plane to probe and characterize objects in the deep Kuiper Belt along the direction of flight of the NASA New Horizons spacecraft. It will probe the deep Kuiper belt as a function of distance and allow us to better understand the streaming instability process of Kuiper Belt object formation and elucidate crater formation physics on these icy bodies.  It will enable KBO rotation studies, in particular of those objects with long periods. Discover and characterize Kuiper belt object binaries, important because their duplicity offers information about object densities at these distant locations from the Sun. It will also shed light on the cratering history of KBOs and dating the surfaces of Arrokoth, Pluto and Charon in addition to helping to place the 32 distant KBOs New Horizons has observed in context. This project has synergy with transit exoplanet studies sensitive to discovery of hot Jupiters and hot Neptunes and can provide important calibration measurements for constraining Galactic models for microlensing studies. 

\bibliographystyle{aasjournal}
\bibliography{NHKBOs_v1}

\end{document}